\documentclass[%
reprint,
amsmath,amssymb,
aps,
prb,
]{revtex4-2}

\usepackage{graphicx}
\usepackage{dcolumn}
\usepackage{bm}
\usepackage{hyperref}

\usepackage{float}
\usepackage{caption}
\usepackage{subcaption}
\usepackage{lipsum}  

\begin{document}

\preprint{APS/123-QED}

\title{Lanes and lattice structures in a repulsive model for self-propelled agents}

\author{P. Bisht}
\email{bishtp4496@gmail.com}
\affiliation{%
U R Rao Satellite Centre, ISRO, HAL Old Airport Road, Vimanapura, Bengaluru, Karnataka 560017, India}%


\begin{abstract}

We investigate a simple Vicsek-type rule-based model for self-propelled particles, where each particle orients itself antiparallel to the average orientation of particles within a defined neighborhood of radius \( R\). The particle orientation is updated asynchronously and randomly across the system. In steady state, particles self-organize into clusters—despite the repulsive interaction—and form two interwoven hexagonal lattices moving in opposite directions chosen spontaneously. Increasing noise in the reorientation step reduces the laning effect, but the global crystalline order remains intact at sufficiently high densities. The mean-squared displacement exhibits super-diffusive growth \( \sim t^{3/2} \) in the transient phase, transitioning to ballistic motion \( \sim t^2 \) in the steady state in the high density and zero noise regime. With an increase in noise and/or decrease in density, the mean-squared displacement grows diffusively \( \sim t \). We observe a cutoff for the ratio \( \frac{R}{L} \sim 0.2-0.3 \), below which laning and crystallization is achieved, suggesting a local but non-microscopic sphere of influence is required to initiate laning.
\end{abstract}

\maketitle


\section*{Introduction}
The defining feature of active matter is the emergence of self-organization and collective motion displayed by its constituent agents \cite{marchetti, bechinger}. From the spontaneous rearrangement of the cytoskeleton initiating cellular motility \cite{cytoskeleton} to the onset of marching in migrating locusts \cite{locusts}, or the unanimously directed motion of a collection of rolling colloids \cite{rollingcolloids}, global coordination and transport generally emerge in these systems solely on the basis of local rules of cooperation. Agent-based models that retain only the basic interaction rules, such as the standard Vicsek model \cite{vicsek}, have been instrumental in understanding the phenomenology of collective behavior like the existence of a wide range of phases and phase transitions, the effect of density, noise, dimension thereon, etc. \cite{tonertu,ginelli}.

Not all collective behaviors stem from cooperative interactions. Conflict, arising from steric effects, volume exclusion, or the necessity to maintain personal space \cite{sciortino2021, fily2012, redner, bialke2013}, often produces distinctive spatiotemporal patterns \cite{sheep, glidingbacteria}. For instance, in human contexts, unidirectional streams spontaneously emerge in a crowd of pedestrians moving in opposite directions—a fascinating solution for optimizing collective transport \cite{karamouzas2014, moussaid, guy}. This phenomenon illustrates individuals’ ability to balance the conflicting demands of coordinating with the crowd while navigating through it.

The \textit{social force model} \cite{sfm} and the \textit{floor-field cellular automaton model} \cite{ffm} are prevalent frameworks for describing pedestrian dynamics. The former focuses on the interaction potential between pedestrians while the latter works with an abstract interaction between pedestrians and an evolving floor underneath. However, traditional implementations of these models often assume instantaneous reactions to stimuli, overlooking the anticipatory nature of human navigation. Increasingly, field studies have demonstrated that mutual anticipation plays a pivotal role in the emergence of structured phenomena such as lane formation \cite{murakami,murakami2,tomaru,cirillo}. This insight has spurred the incorporation of anticipatory mechanisms in existing models to better capture these dynamics \cite{suma2012, bailo2019, cirillo}.

In this study, we propose a minimal rule-based model for self-propelled agents with repulsive interactions that reproduces the laning phenomena. In this model, each particle reorients to the polar opposite of the average orientation of its neighbors within a radius R. Reorientation updates occur asynchronously to approximate the delays inherent in real-world interactions, such as those observed in systems governed by anticipation and reaction times.

In the steady state, the agents spontaneously segregate into two distinct groups moving in opposite directions, forming parallel lanes. Within each lane, particles group into evenly spaced clusters—a counter-intuitive outcome given their repulsive interactions. These clusters organize into interwoven hexagonal lattices, with the direction of motion emerging spontaneously, breaking rotational symmetry. Signatures of clustering and crystallization have been reported before in systems governed by \textit{anticipatory} potential \cite{gerlee2017}. 
Self-organization is influenced by the interplay of noise and density, similarly to what is observed in the Vicsek model. Higher levels of noise or lower particle density disrupt laning. Despite this, high-density systems are able to maintain global crystalline order even under high degrees of noise, similar to a phenomenon earlier reported as \textit{freezing by heating}\cite{freeze_by_heat}.

    \begin{figure*}[ht!]
        \begin{subfigure}{.25\textwidth}
            \centering
            \includegraphics[scale=0.175]{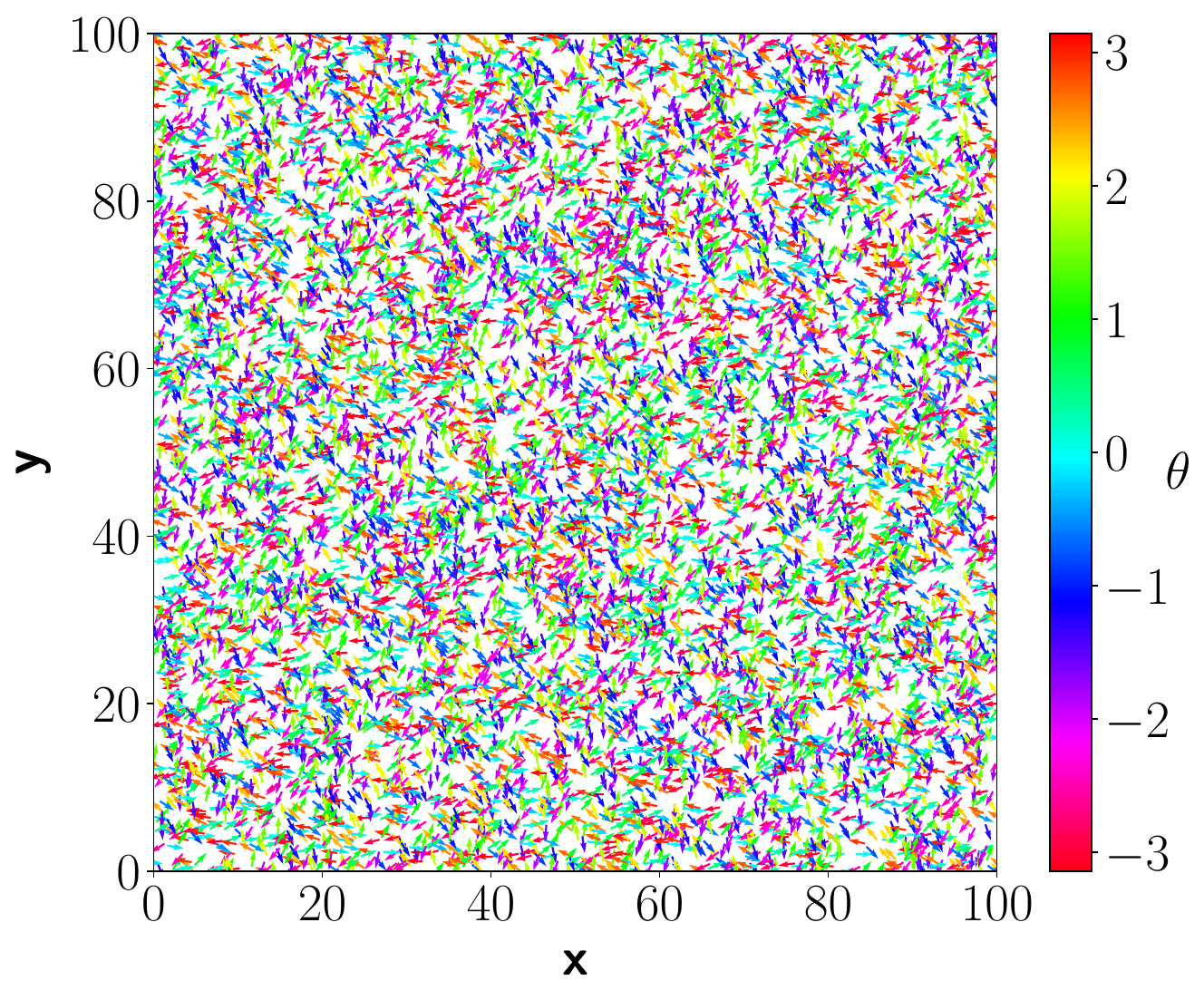}  
            \caption{}
        \end{subfigure}
        \begin{subfigure}{.2\textwidth}
            \centering
            \includegraphics[scale=0.2]{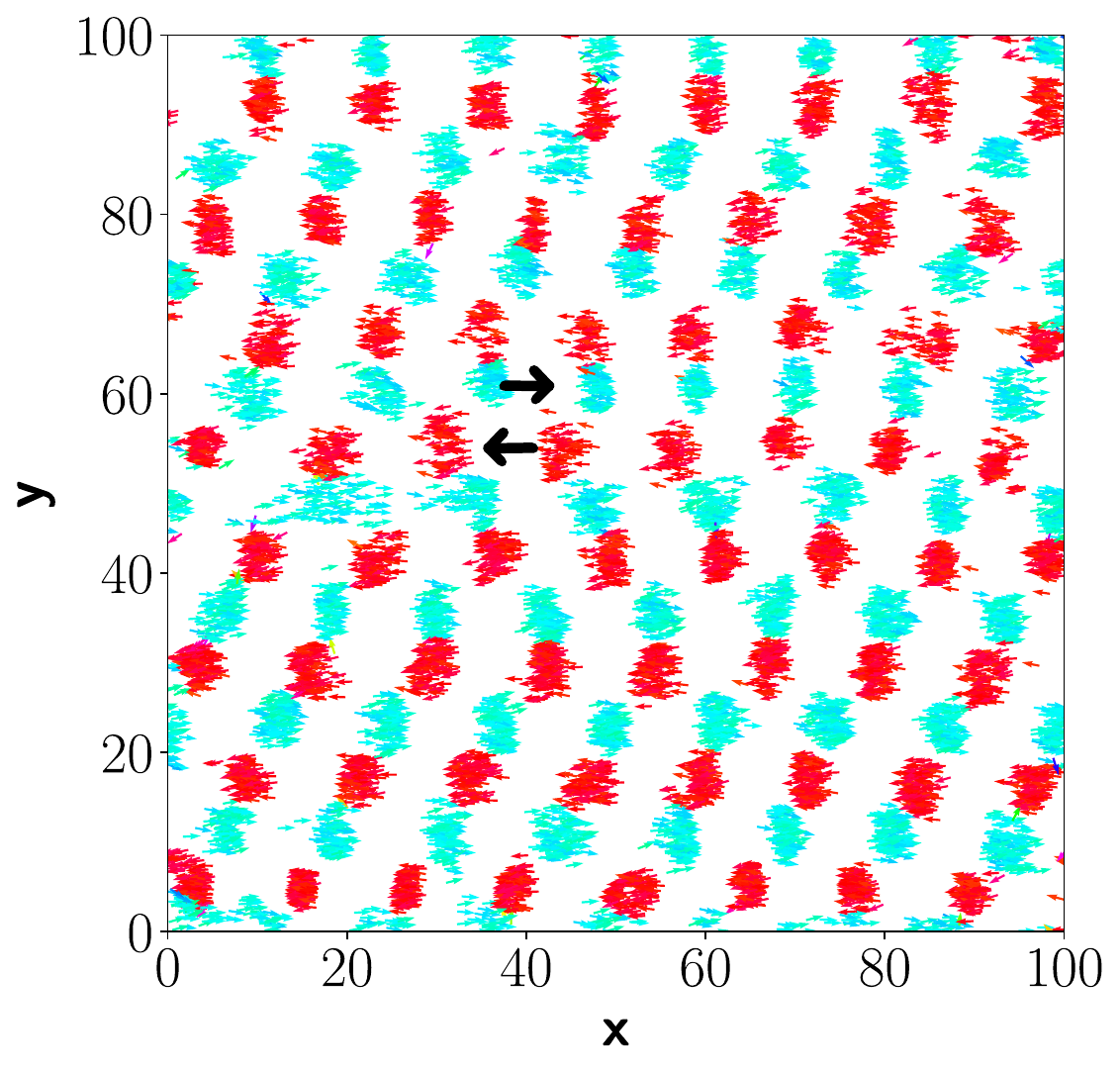}  
            \caption{}
        \end{subfigure}
        \begin{subfigure}{.2\textwidth}
            \centering
            \includegraphics[scale=0.2]{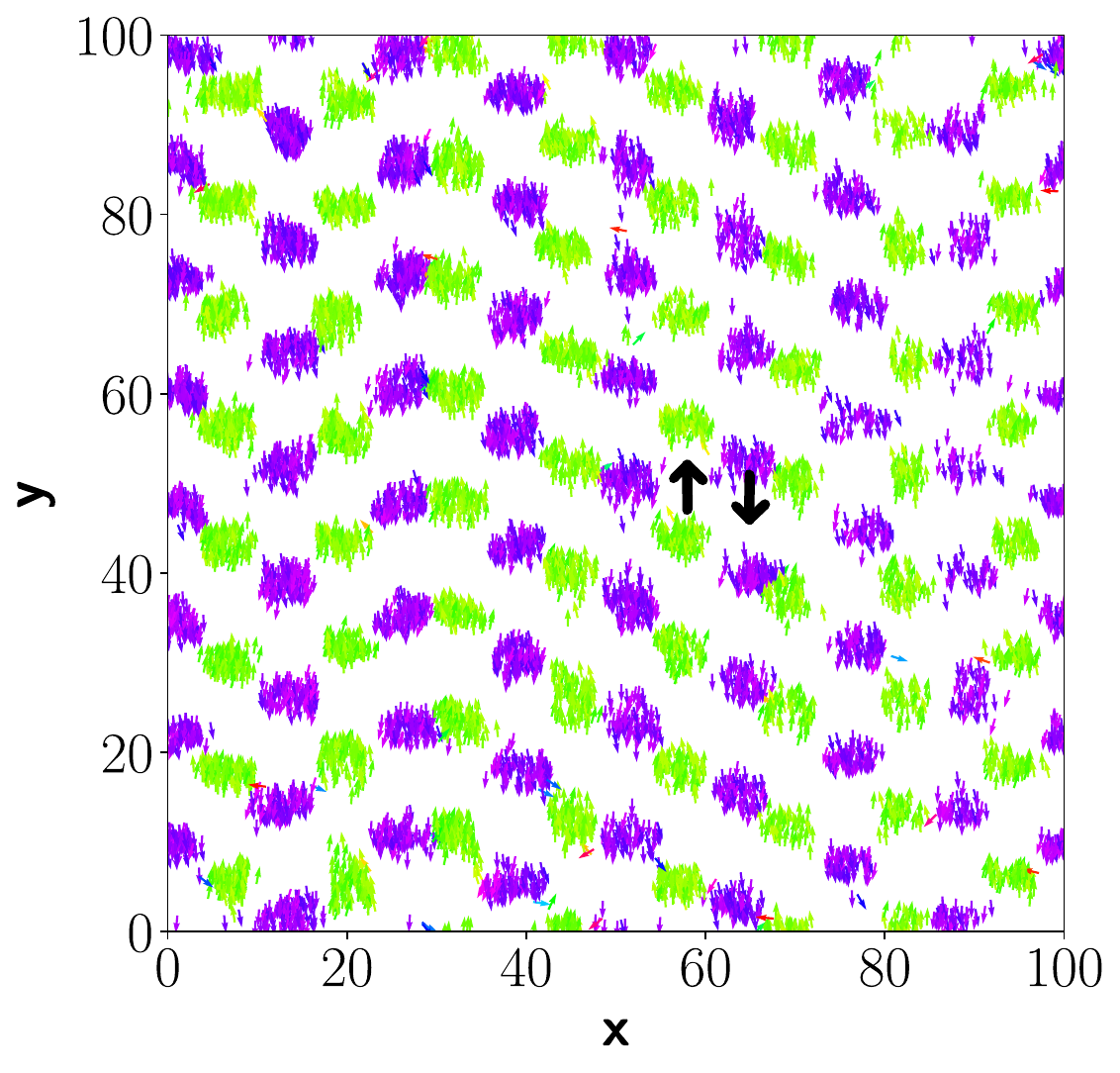}
            \caption{}
        \end{subfigure}
        \begin{subfigure}{.2\textwidth}
            \centering
            \includegraphics[scale=0.2]{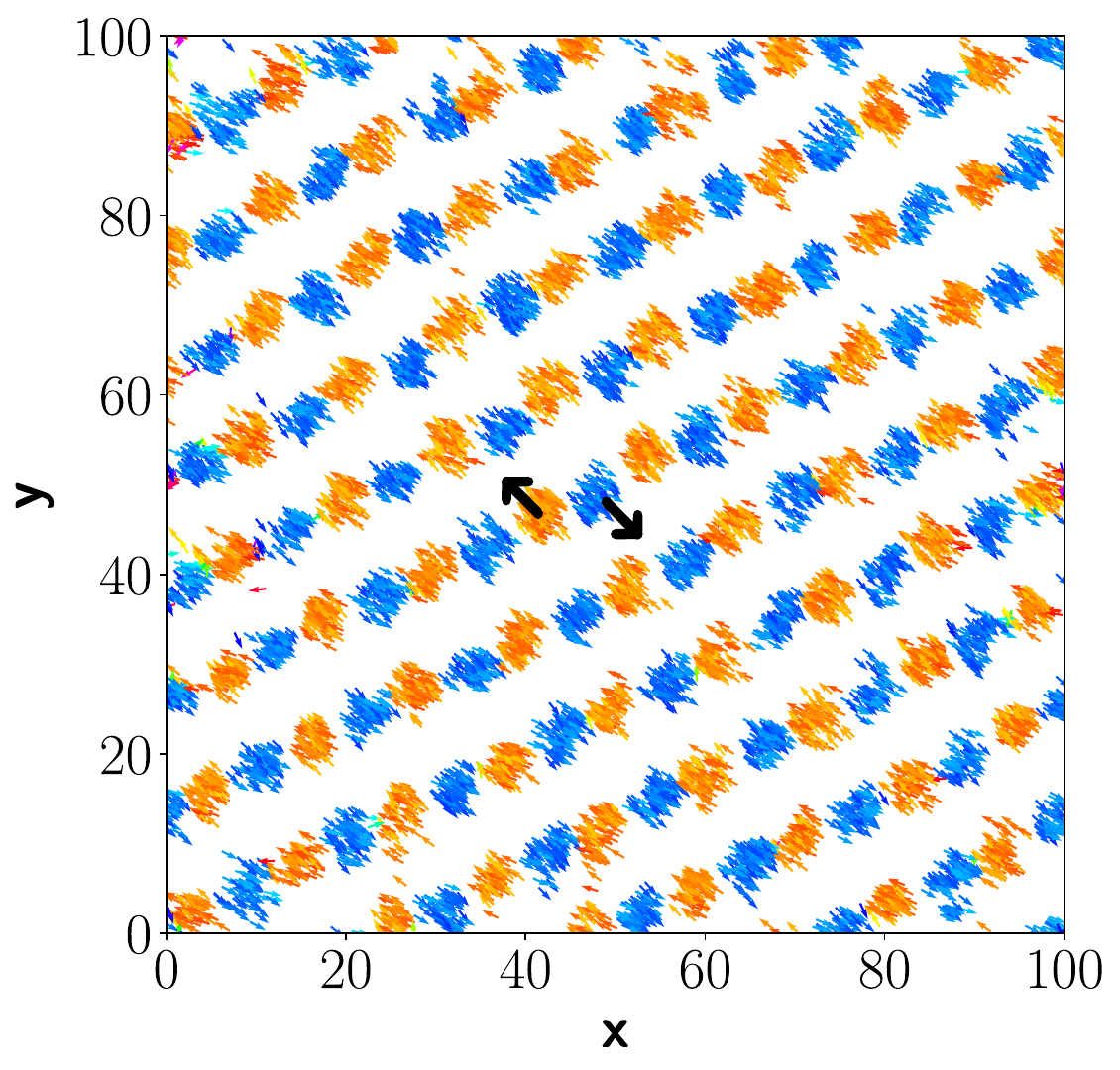}   
            \caption{} 
        \end{subfigure}
        
        \caption{(a) A typical initial state of the system; (b), (c), (d) steady-state snapshots for 3 simulations of a system with $L=100, N=10,000, R=0.1L$ and $\eta = 0.5$ at the simulation time $T=5000$. The colour of a particle indicates its orientation and hence the direction of its motion (black arrows), per the colorbar in (a). Plots have 16$\times$8 roughly equally distributed clusters.}
        \label{fig:sss}
    \end{figure*}

The primary aim of this study is to investigate the organizational features of a repulsive and delay-guided active systems using a minimal algorithmic model, akin to the role Vicsek model plays in flocking behavior. This approach may complement modeling approaches in non-equilibrium systems where antagonistic interactions facilitate crystallization among agents with non-zero velocities \cite{bialke, caprini, sanchez-miranda}, a subject of growing interest in the field.  Despite its simplicity, our model captures two distinct phases observed in repulsive multiparticle systems: laning and \textit{freezing-by-heating}-type crystallization \cite{freeze_by_heat, freeze_by_heat2} (which are organizationally distinct from the well-known motility-induced phase separation (MIPS) seen in other repulsive active matter systems \cite{palacci, cates}). We propose that these two phenomena originate from the same underlying dynamics, with their manifestation determined by varying noise levels. 

Furthermore, our findings reveal that reflecting boundary conditions are not essential for laning or crystallization, as previously suggested, and that particles do not need to initially move in opposite directions to display these behaviors \cite{leunissen, rex, menzel2, ribeiro}.  Instead, the asynchronous update rule plays the most important role, second to repulsion, in enabling the onset of these phases; under synchronous update mechanism, particles just exhibit the classic Vicsek-type flocking behavior.

\section*{Model}
We take a collection of $N$ point particles with randomly initialized positions $(x,y)$ in a 2D box of size $L\times L$ with periodic boundary conditions and orientation $(\theta)$ randomly assigned from the interval $-\pi$ to $\pi$. In one time-step $\Delta t$, the particle position is incremented as:

\begin{equation}
\begin{bmatrix}
\Delta x_i(t)\\
\Delta y_i(t)
\end{bmatrix}  
= v_o 
\begin{bmatrix}
\cos{\theta_i(t)}\\
\sin{\theta_i(t)}
\end{bmatrix}  
\Delta t
\end{equation}

where $v_o$ is the velocity magnitude and $\Delta t$ signifies one time step. Both the parameters are constant and set to $v_o = 1$ and $\Delta t = 1$ in all simulations. Following this, the orientation for the particles is updated randomly and in sequential fashion, as per the rule

\begin{equation}
\theta_i(t+1) = \langle\theta(t)\rangle_R  + \Delta \theta_i
\label{eq:update1}
\end{equation}

where $\langle\theta(t)\rangle_R$ is the mean orientation of the neighbourhood around the $i^{th}$ particle, restricted to a ball of radius $R$:

\begin{equation}
\langle\theta(t)\rangle_{R} = \arctan\Bigg[\frac{\langle\sin{(\theta(t) + \xi)}\rangle_R}{\langle\cos{(\theta(t) + \xi)}\rangle_R}\Bigg]
\label{eq:update2}
\end{equation}

here we set $\xi = \pi$ indicating that the particle assumes an orientation opposite to the mean orientation of its neighborhood instead of aligning parallel to it, for which $\xi = 0$ (as for the Vicsek model). The term $\Delta \theta_i$ is a random number chosen from $[-\frac{\eta}{2}:\frac{\eta}{2}]$ with uniform probability, where $\eta$ signifies the degree of noise. The orientation update step for the particles is done in the sequence of the particle order, which is reshuffled at the end of one time step. The algorithm for the model can be summarized as:

{\small{
    \begin{enumerate}
        \item Initialize position and orientation for $N$ particles. The positions are uniformly distributed in a $L\times L$ box  with periodic boundary conditions, and the orientations are chosen from the uniform distribution in [$-\pi$:$\pi$].
        \item In one time step:
        \begin{itemize}
            \item Based on current orientation and hence velocity vector, update all particle positions.
            \item Based on Eq. \ref{eq:update1},\ref{eq:update2}, update particle orientation in random order, sequentially.
        \end{itemize}
        \item Repeat step 2 for $T$ number of time-steps.
    \end{enumerate}
}}

Like in the classic Vicsek model, our particles follow non-exclusion at any given site.
\section*{Results}

The plots in Fig. \ref{fig:sss}(b,c,d) show steady state snapshots of the system at the end of three independent simulations starting from a randomly initialized state like one shown in Fig. \ref{fig:sss}a for a system with $N=10,000$ particles, $L=100$ system size ($\rho = N/L^2 = 1$), radius of interaction, $R=0.1L$ and magnitude of noise $\eta$ set to 0.5. The particles aggregate in clusters, which further arrange themselves into regularly spaced units in a ''lane." Overall, there are two such lanes moving alternately in opposite directions. 

\begin{figure}[H]
    \centering
    \includegraphics[width=\linewidth]{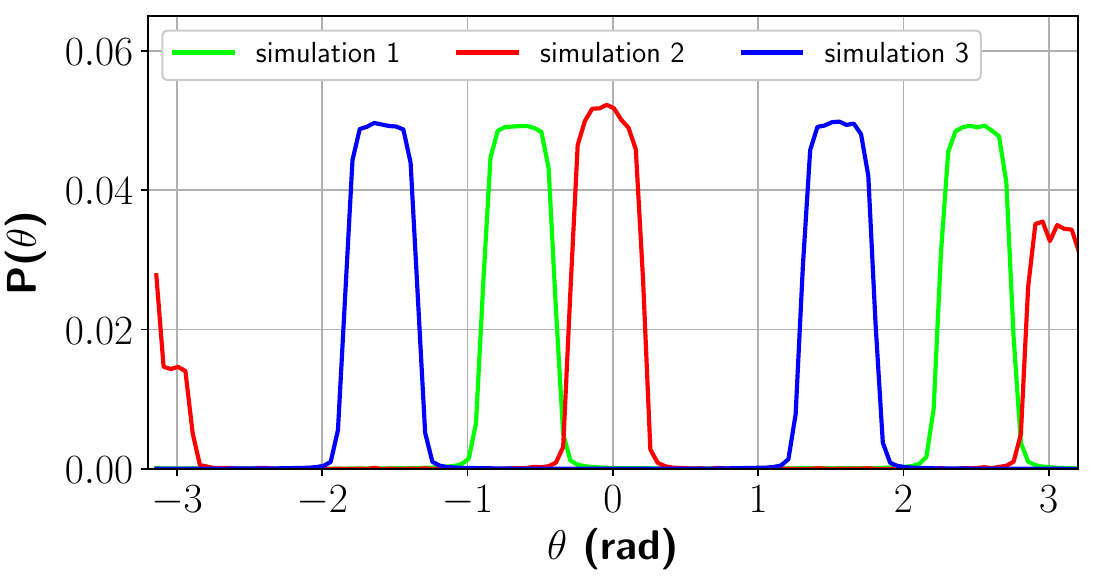}
    \caption{Probability distribution of orientation $(\theta: [-\pi, \pi])$ of particles in steady state for the simulations in Fig. \ref{fig:sss}(b,c,d) respectively.}
    \label{fig:pdftheta}
\end{figure}

Fig. \ref{fig:pdftheta} is a histogram showing the distribution of orientation $\theta$ of $N$ particles ($\theta: [-\pi:\pi]$) in steady state for the said simulations. The histogram shows there are two peaks $\{\theta_{max}, \theta_{max} + \pi\}$ separated by a value of $\pi$ in a given simulation. The value of $\{\theta_{max}\}$ is random and unpredictable for a given simulation, i.e., the direction of this bi-lane system in the steady-state is selected spontaneously and remains so throughout (see Fig. \ref{fig:evolution}). 

\begin{figure}[H]
    \centering
    \includegraphics[width=\linewidth]{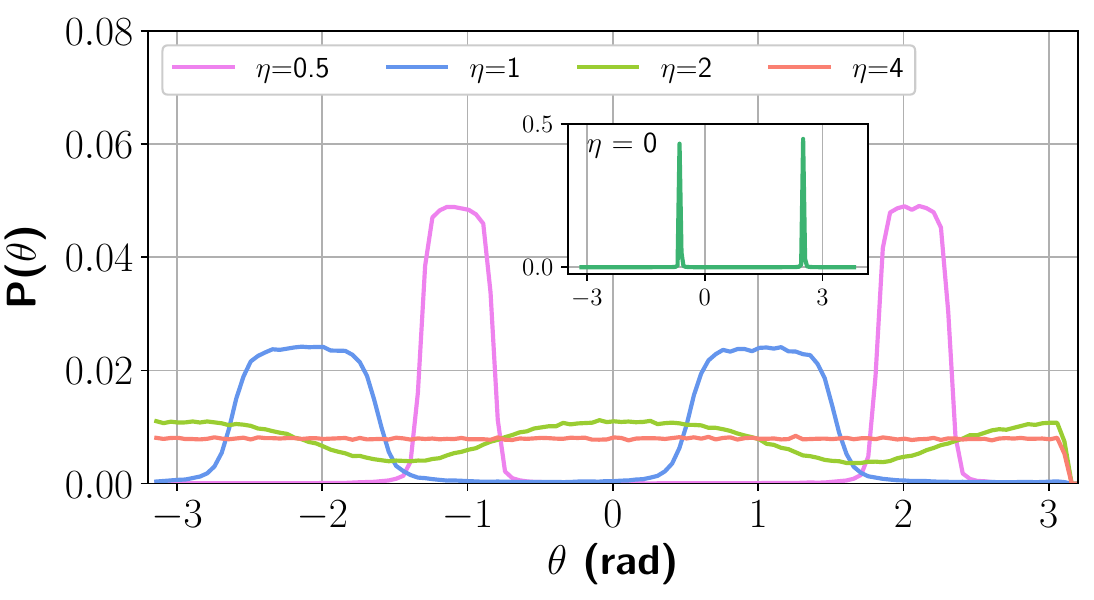}
    \caption{$P(\theta)$ in steady state for different levels of noise. Inset shows $P(\theta)$ for $\eta=0$ case.}
    \label{fig:pdfthetaeta}
\end{figure}

Figure \ref{fig:pdfthetaeta} shows $P(\theta)$ for five different values of $\eta$. 

\begin{figure}[!h]
    \begin{subfigure}{.225\textwidth}
        \centering
        \includegraphics[trim={0 0 3.1cm 0},clip, scale=0.25]{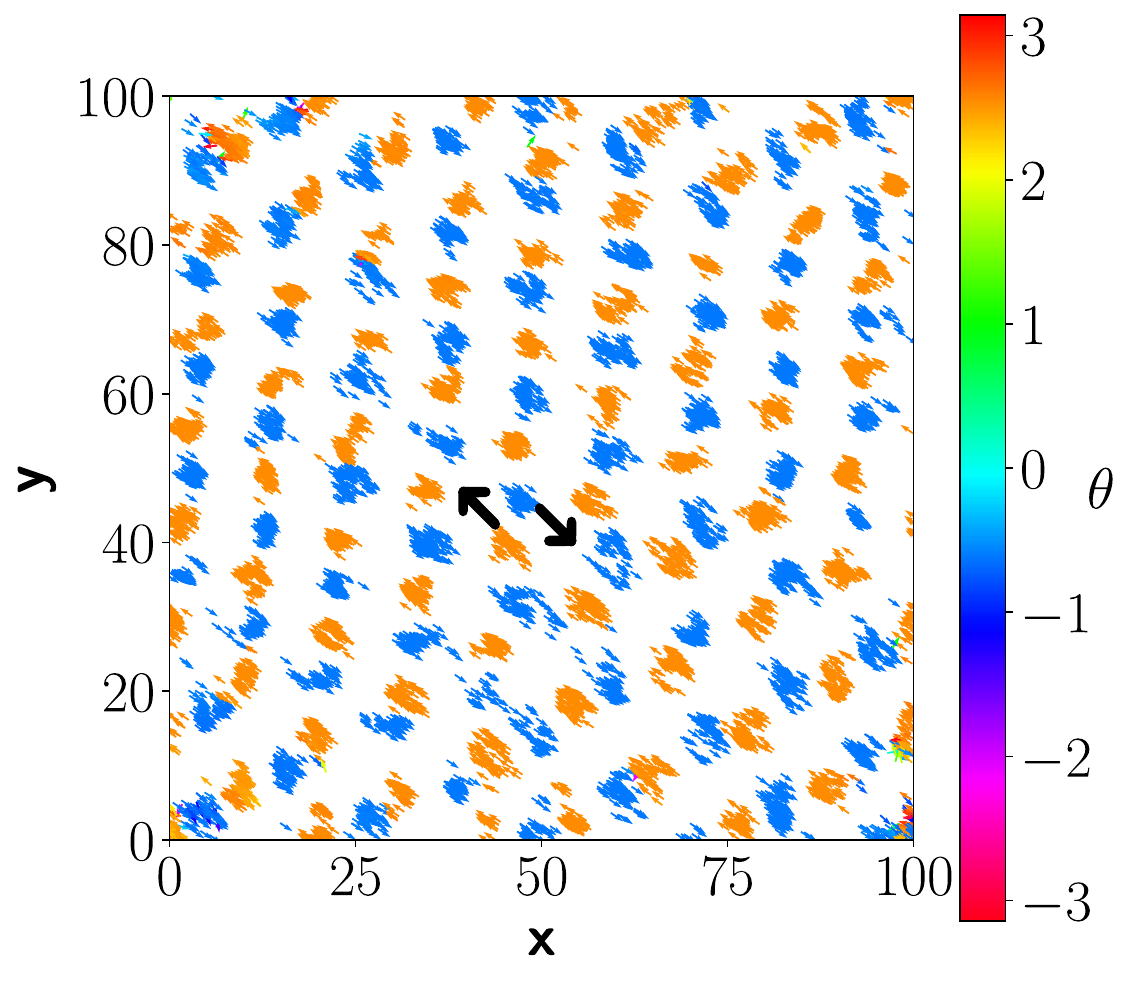}  
        \caption{}
    \end{subfigure}
    \begin{subfigure}{.225\textwidth}
        \centering
        \includegraphics[scale=0.25]{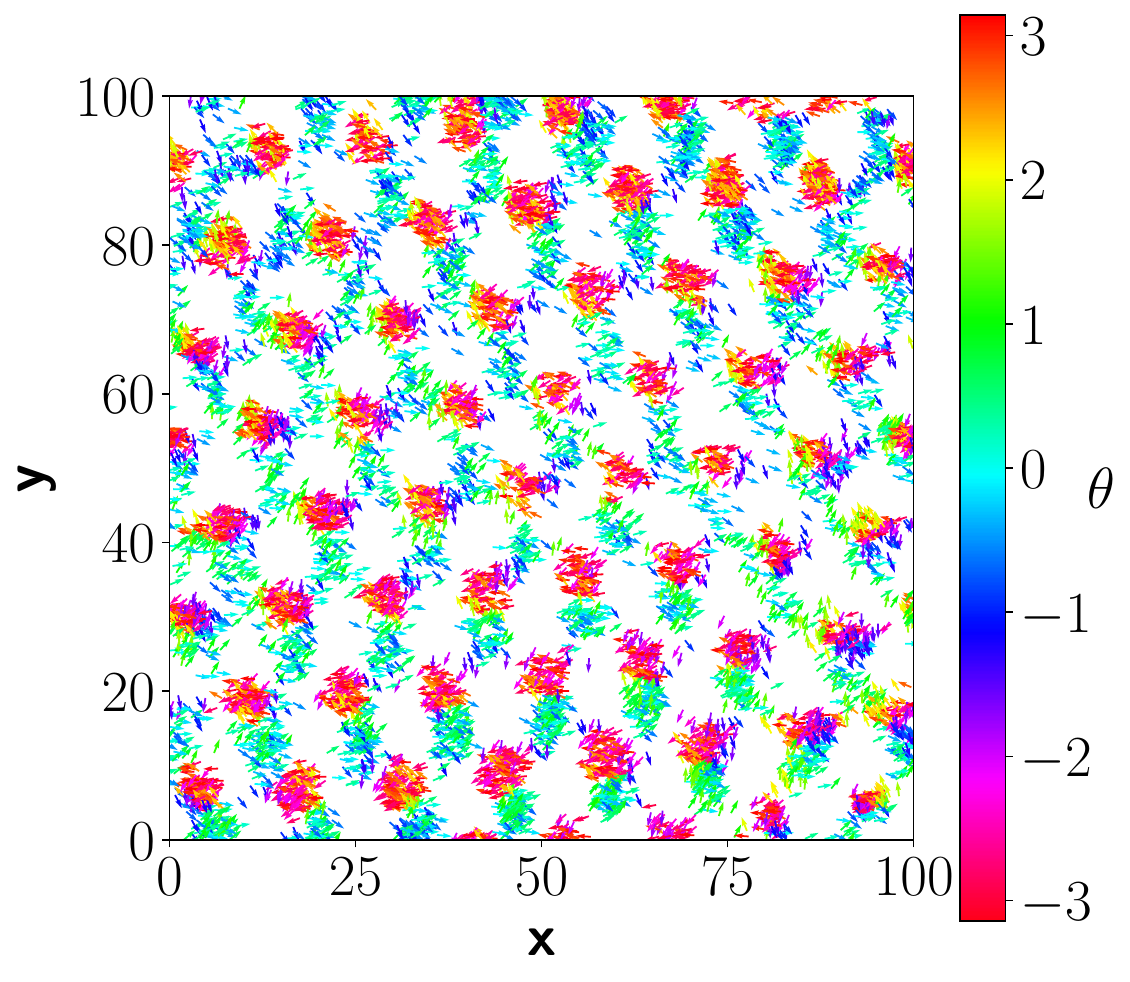}  
        \caption{}
    \end{subfigure}
    \newline
    \begin{subfigure}{.225\textwidth}
        \centering
        \includegraphics[scale=0.21]{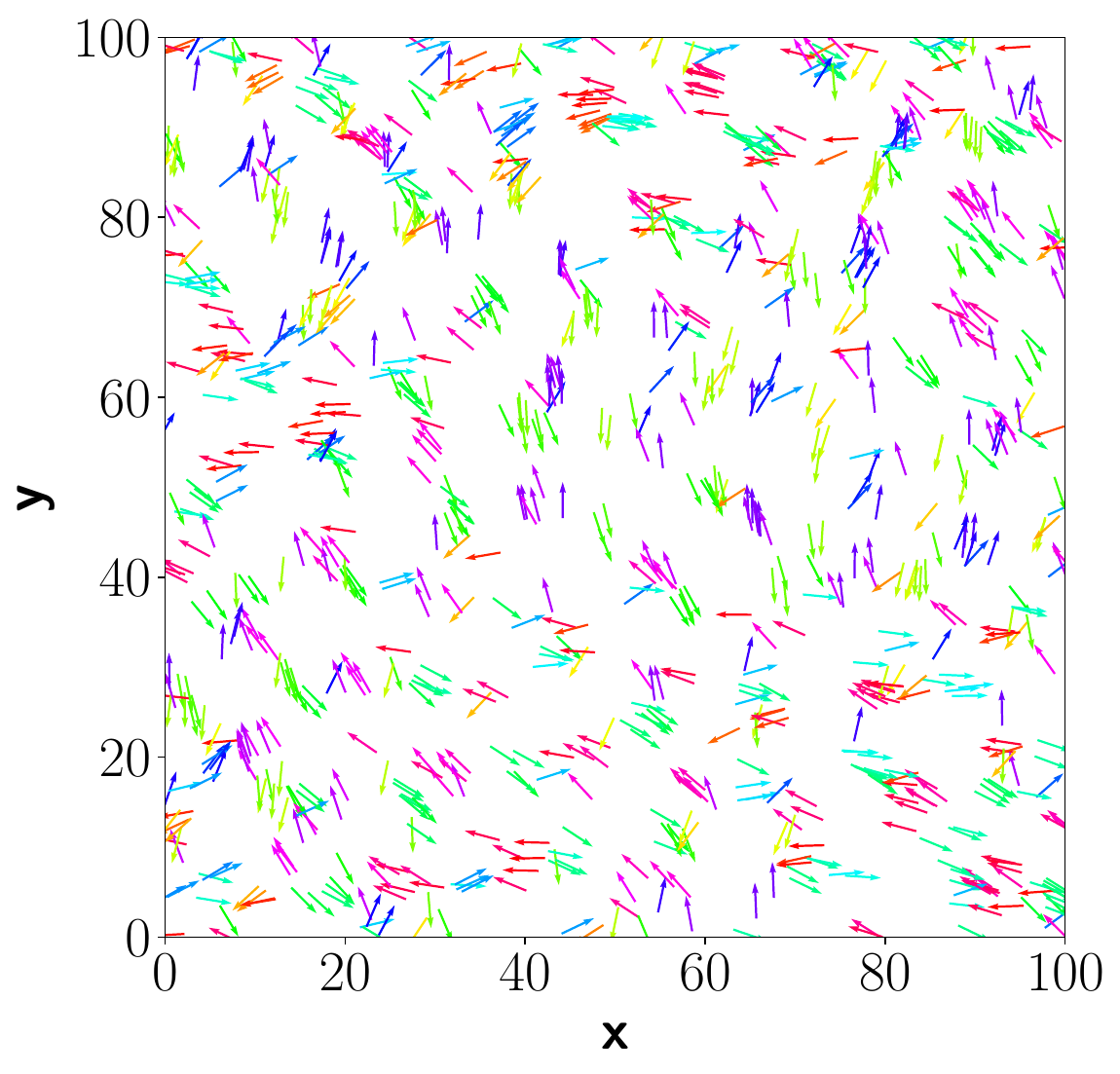}  
        \caption{}
    \end{subfigure}
    \begin{subfigure}{.225\textwidth}
        \centering
        \includegraphics[scale=0.21]{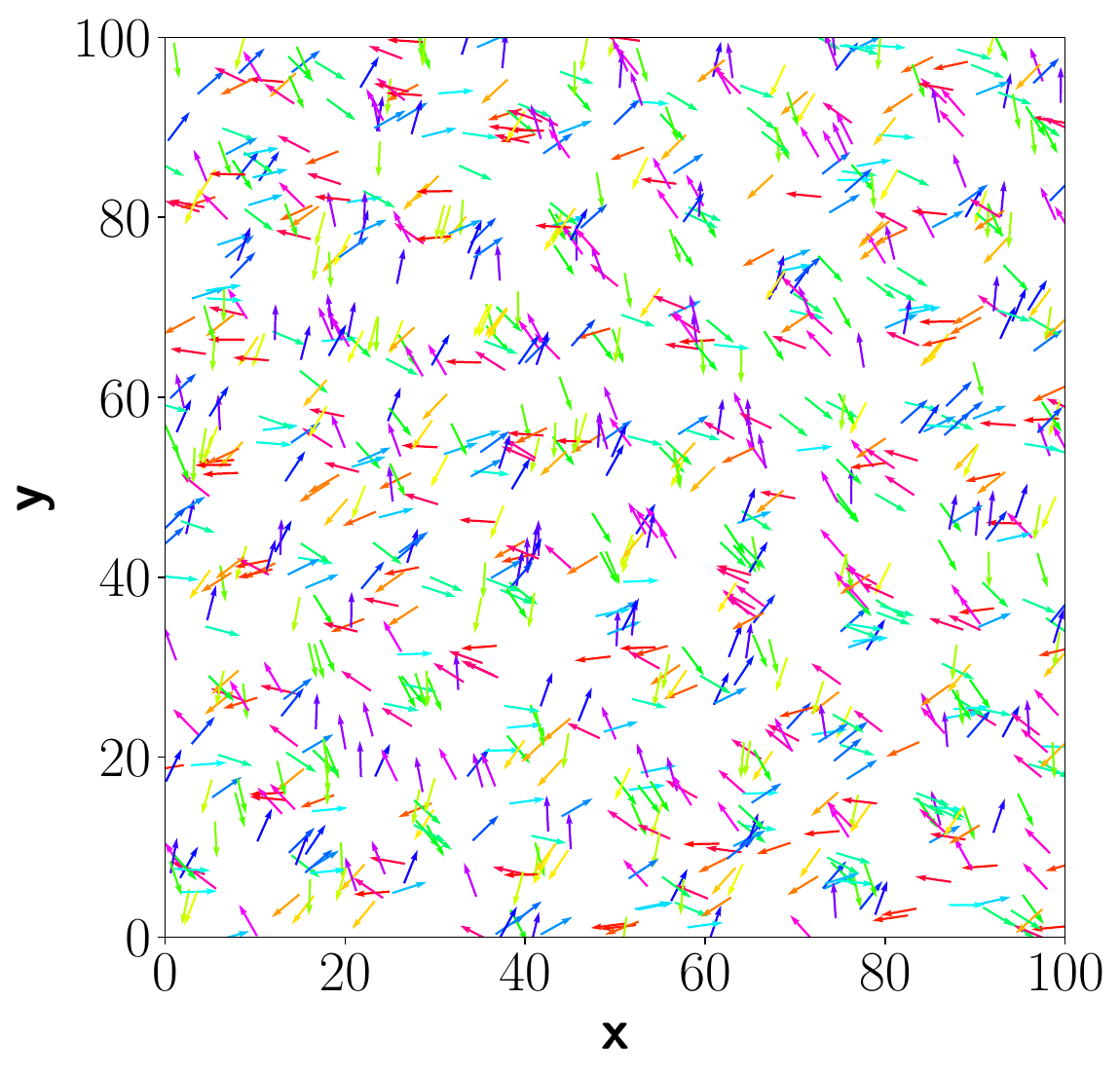}  
        \caption{}
    \end{subfigure}
    
    \caption{(a) Steady-state snapshots for systems with varying density and noise. For each case, the system size is fixed to $L=100$ and $R=0.1L$. (a) $\rho = 1$, $\eta = 0$, high density, low noise case. This is the ordered phase where particle clusters are embedded in migrating lanes of two opposing polarities. (b) $\rho = 1$, $\eta = 2$, high density, high noise. Orientational order tends to vanish 
    but there is some degree of spatial order. (c) $\rho = 0.1$, $\eta = 0$, low density, low noise. Local particle clusters are observed but there is no global orientational order. (d) $\rho = 0.1$, $\eta = 2$, low density, high noise. No orientational or spatial order exists.}
    \label{fig:noisedensityeffect_vis}
\end{figure}

With increasing levels of $\eta$, particle distribution becomes more and more dispersed about the chosen axis of motion (the fwhm varies as $\sim \eta$), until eventually all polarity gets lost and particle orientation becomes uniformly distributed globally. Figure \ref{fig:noisedensityeffect_vis}(b) shows how noise, coupled with the repulsive interaction rule, precludes organization and directed motion; the particles tend to simply fluctuate and diffuse around their initial position as the noise level increases. It is interesting to note that even for higher levels of disorder, $\eta=2$ here ($\eta$ cannot exceed $\pi$), there is an overarching spatial order in particle density. A similar phase with crystalline order was observed in a repulsively interacting multiparticle system sustained by a high degree of total energy/fluctuations and was paradoxically named \textit{freezing by heating} \cite{freeze_by_heat}.

\begin{figure}[!h]
    \centering
    \includegraphics[width=\linewidth]{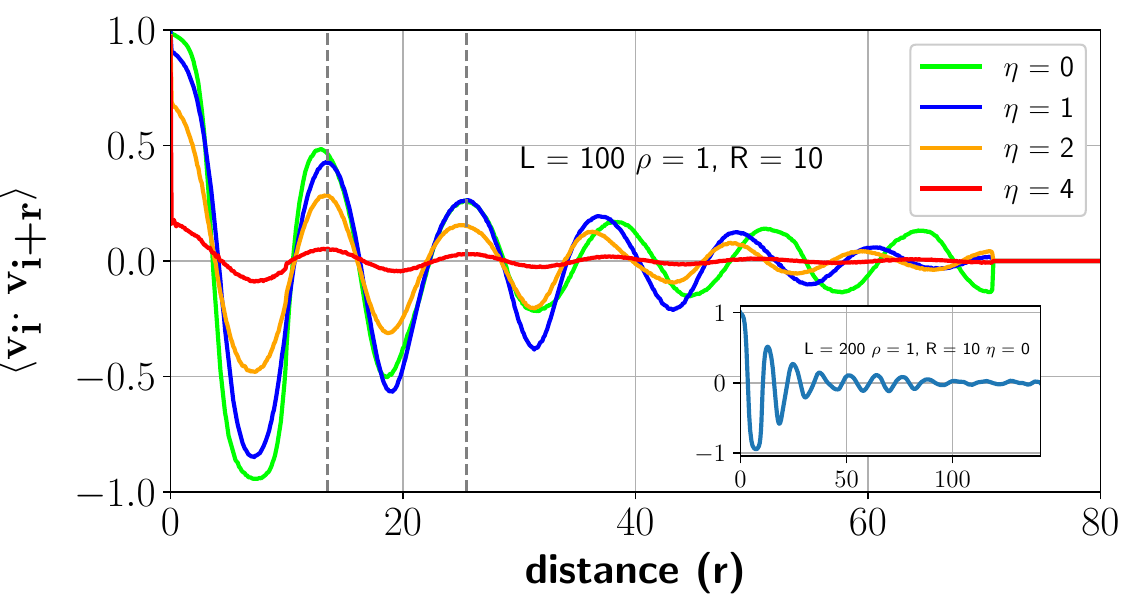}
    \caption{Average two-point correlation function for particle velocity in steady state for different levels of noise. 
    }
    \label{fig:2pvcR}
\end{figure}

Figure \ref{fig:2pvcR} shows the average equal-time two-point velocity correlation $C_v(r) = \langle v_i \cdot v_{i+r} \rangle $ of particles in the steady state with increasing levels of noise. 
The cutoff in the function, for $\eta=0$ case is an effect of the finiteness of the system size under consideration (since  $r_{\textrm{max}} = \sqrt{2}L/2$). 

In a larger system, the velocity correlation at \(\eta = 0\) decays to zero as the separation \(r \rightarrow O(L)\) (see Fig. \ref{fig:2pvcR}, inset). The shape of this function is reminiscent of the two-point spin correlation in an antiferromagnetic material, where the role of temperature is replaced by noise in our model. Once again, even in the presence of high disorder (\(\eta = 2\)), the velocity correlation function suggests the existence of crystalline spatial order on the scale of the system size. This is consistent with recent findings where the radial-correlation function of self-propelled particles, interacting solely via repulsive torques, demonstrated crystallization under conditions of high density and strong repulsion (or, low speeds) \cite{leblay}.

\begin{figure}[!h]
    \centering
    \includegraphics[scale=0.33]{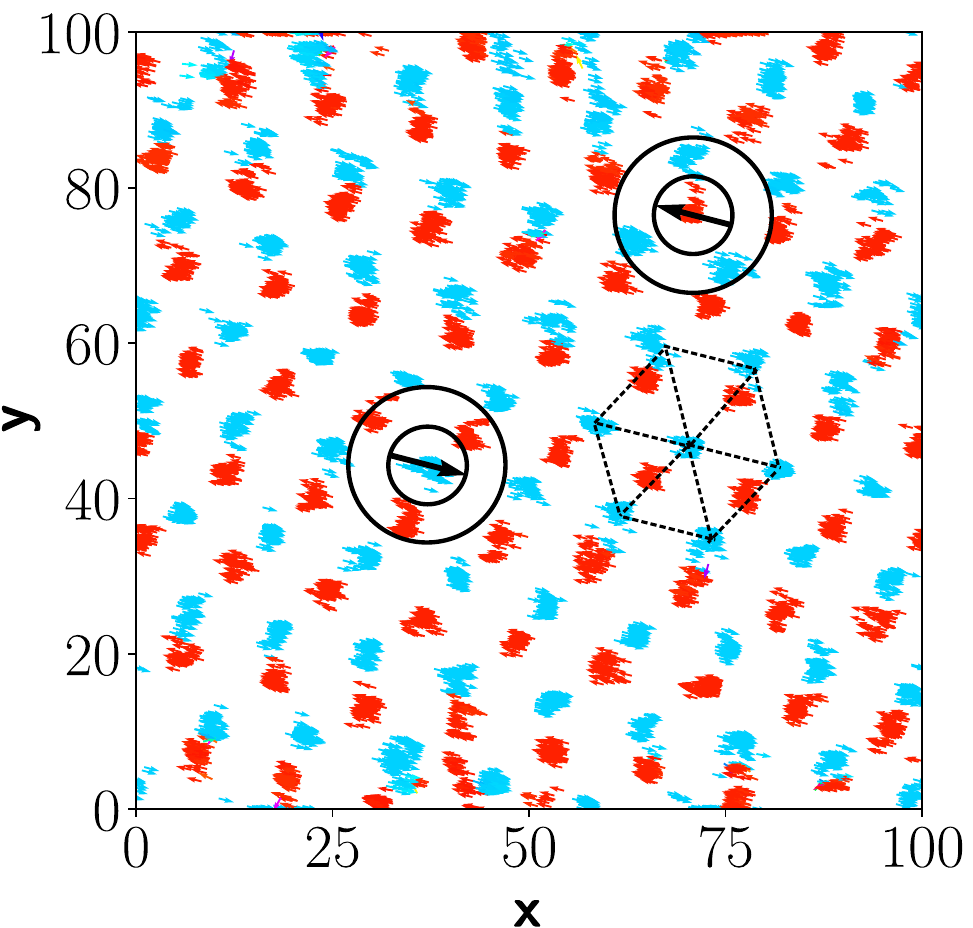}  
    \caption{Steady-state snapshot($L=100$, $\rho=1, R=0.1L, \eta=0$) marking tagged particles by a black arrow, indicating its direction of motion. The radii of the inner and outer circles are $R/2$ and $R$ respectively. Dotted lines trace out a hexagonal lattice made of clusters of the same polarity. 
        }
    \label{fig:ss}
\end{figure}

The scenarios in Figure \ref{fig:noisedensityeffect_vis}(c,d) correspond to low-density counterparts of (a,b), respectively. The steady state in (c) shows local spatial clustering of particles (the global order is lost), while the clusters themselves follow a gaseous diffusive motion. Meanwhile, in (d), both orientational and spatial order is lost. We evaluate the mean-squared displacement (MSD) plot of the particles under all scenarios and detect the presence of any caging effect in particle motion. We define MSD as:

\begin{equation}
\textrm{MSD} = \left< r^2(t) \right> = \frac{1}{N}\sum_{i=0}^{N} |r(t+t_0) - r(t_0)|^2 
\end{equation}

where $\left< \right>$ denotes the average over all particles and $t_0$ denotes the time at the beginning of the simulation (periodic boundary is not applied while calculating MSD \cite{v-lee}).

\begin{figure}[!h]
    \centering 
    \includegraphics[width=\linewidth]{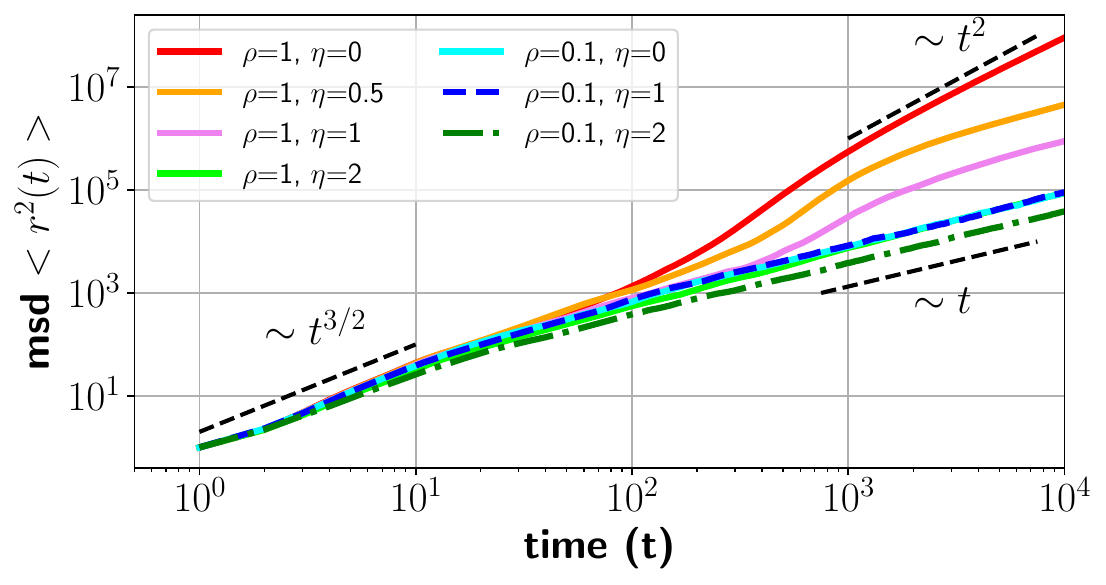}
    \caption{Mean-squared displacement of a tagged particle for different cases in Fig. \ref{fig:noisedensityeffect_vis}.}
    \label{fig:msd}
\end{figure}

For the case \(\rho = 1, \eta = 0\), we initially observe a superdiffusive transient regime where \(\left< r^2(t) \right>\) grows as \(\sim t^{3/2}\) (see Fig. \ref{fig:msd}). After a time \(\tau\), which corresponds to the onset of the steady state, the mean-squared displacement (MSD) grows ballistically, i.e., \(\left< r^2(t) \right> \sim t^{2}\), consistent with the lock-step directed motion of the particles in steady state. This suggests that particles, as a whole, do not switch lanes or clusters in the steady state.

For all other values of \(\eta > 0\), the superdiffusive transient regime transitions into a diffusive displacement growth, i.e., \(\left< r^2(t) \right> \sim t\). In this case, the motion of particles resembles a random walk; displacements are driven by random fluctuations rather than directed behavior, and the diffusion coefficient decreases with increasing noise. Although there is a global spatial order (Fig. \ref{fig:noisedensityeffect_vis}.b), particles are migrating diffusively and hopping between clusters in the process.

In the low density regime \(\rho = 0.1\), we observe a diffusive displacement growth in the steady state for all values of $\eta$. A recent study on the emergence of crystalline phases in a 2D magnetic granular system with short-range repulsive interactions reported similar trends in MSD evolution across phases with varying particle concentrations \cite{sanchez-miranda}.

\begin{figure}[!h]
    \centering
    \includegraphics[width=\linewidth]{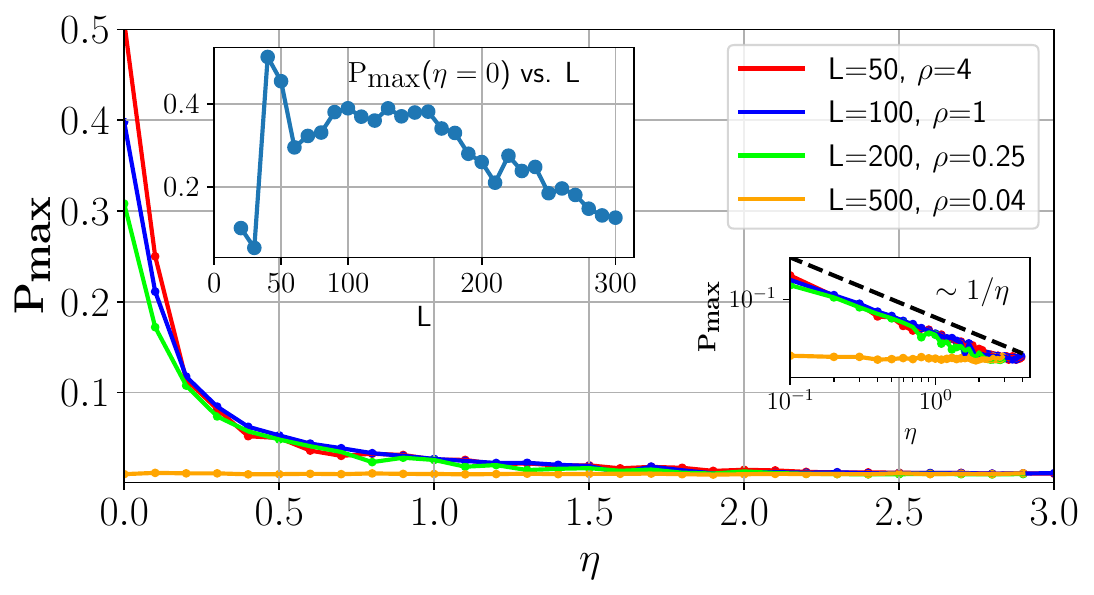}
    \caption{$P_{\textrm{max}}$ i.e., $\rm{max}_{\theta \in [-\pi:\pi]}\{P (\theta) \}$ with respect to noise strength $\eta$ for $N=10,000, R=10$ and varying system size $L$. The inset at the right shows the same plot in log scale. The inset at the top shows $P_{\textrm{max}}(\eta = 0)$ with respect to $L$ for fixed $R=10$.
    }
    \label{fig:noiseeffect}
\end{figure}

In Fig. \ref{fig:noiseeffect} we observe the polarity in steady state, represented by the quantity, $\rm{max}_{\theta  \in [-\pi:\pi]}\{P (\theta) \}$ i.e. $P_{\textrm{max}}$, the maxima of the orientation distribution function, as a function of noise level $\eta$ for different system sizes $L$. $P_{\textrm{max}}$ can take a maximum value of 0.5 here for $\eta \rightarrow 0$ and $\rho \rightarrow \infty$. $P_{\textrm{max}}$ starts vanishing for around $\eta \gtrsim 1$, for small $L$.  As $L$ increases, or equivalently, particle density $\rho$ decreases, the polarity starts vanishing even at $\eta=0$. The effect of increasing the system size is best judged at noise $\eta \rightarrow 0$. The inset figure in Fig. \ref{fig:noiseeffect} shows the non-monotonic behaviour of $P_{\textrm{max}}(\eta=0)$ with increase in system size, $L$ keeping  all other parameters fixed constant ($R$ and $N$). 

\begin{figure}[!h]
        \centering
        \includegraphics[scale=0.6]{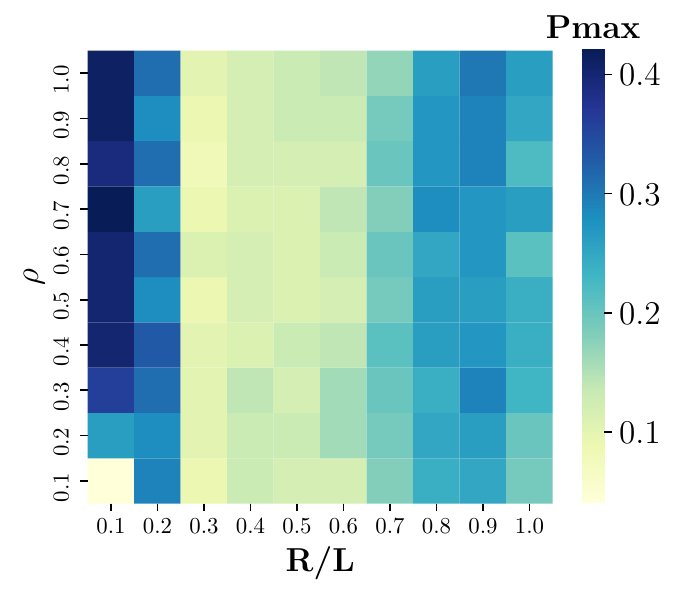}
    \caption{Phase diagram for order parameter $P_{\textrm{max}}$ with respect to particle density $\rho$ and $r=\frac{R}{L}$ at $\eta=0$.
    }
    \label{fig:RbyL}
\end{figure}

So far, we have fixed the radius of influence $R$ to an arbitrary value or a fixed constant while studying the effect of noise, density, system size, etc. Here, we examine the role of the ratio $r=\frac{R}{L}$ in our order parameter $P_{\textrm{max}}$, all at $\eta=0$. In Fig. \ref{fig:RbyL}, we see that, irrespective of the particle density, there is a drop in $P_{\textrm{max}}$ at $r\sim0.2-0.3$. Below this cutoff ($r<0.3$), a strong presence of laning is indicated by the blue region while above ($r<0.3$) laning disappears even at zero noise levels.

The light blue region for $r>0.6$ corresponds to trivial organization, where the entire system falls within a single radius of influence and should be ignored. The effect of $r$ on the order parameter is not yet fully understood, and the possibility of a phase transition remains an object of future investigation. The phase diagram in Fig. \ref{fig:RbyL} suggests that macroscopic neighborhood of interaction, is important to laning and organization. This is evident in the $r=0.1$ case where the order parameter tends to zero. A detailed exploration of the region $r\sim0.1-0.3, \rho\sim 0.1-0.3$ is the subject of future study.

%





\section*{Discussion and Conclusion}
In conclusion, we have studied a simple, repulsive adaptation of the standard Vicsek model, where each particle assumes an orientation polar opposite to the mean orientation of its neighbors. The orientation updates for the entire system are executed in a random, sequential manner, and the radius of influence \( R \), which determines the local neighborhood for a particle, is a finite fraction of the system size.

In the steady state, particles self-organize into two oppositely polarized groups, forming parallel lanes with spontaneously chosen directions. At higher densities, particles cluster into hexagonal arrangements along these lanes. Increasing noise disrupts this order, leading to random motion within a radius \(\sim R\). Lower densities weaken laning and clustering, resulting in dispersed distributions. Noise and density influence the order in a manner similar to the Vicsek model, with higher noise or lower density preventing global organization. This is reflected in the growth of the mean-squared displacement (MSD) of a tagged particle. The MSD shows super-diffusive growth \(\sim t^{3/2}\) in the transient phase, transitioning to ballistic motion \(\sim t^2\) in the steady state for the high-density, low-noise case. The MSD grows diffusively \(\sim t\) as \(\eta\) increases and \(\rho\) decreases.

Even at higher levels of disorder (\(\eta = 2\), where \(\eta \leq \pi\)), an overarching spatial order in particle density persists. This resembles the paradoxical \textit{freezing by heating} phase observed in repulsively interacting systems with strong fluctuations.

The sequential procedure for updating particle orientation introduces relative inertia among particles during the reorientation process, mirroring effects commonly observed in natural collective systems \cite{strombom}. Recent research has demonstrated that inertial active particles can undergo crystallization at sufficiently high densities and levels of self-propulsion \cite{liao} with side-effects like anisotropic random walks and delayed reorientation \cite{sprenger2023inertial}.


It is worth noting that some degree of translational crystalline order invariably emerges in systems of active particles interacting via repulsion, such as through pairwise repulsive potentials, steric hindrance, or volume exclusion \cite{huang, leblay, omar}, provided that the activity and particle density are sufficiently high. In our model, the stationary state is sustained by the excluded volume around particle clusters, similar to what is observed in traditional equilibrium crystalline solids. The key factor here is the balance between the activity of the particles and the strength of their repulsive interactions. The activity drives the particles to explore their environment, while the repulsive interactions create exclusion zones. As the particles move and repel each other, they may spontaneously organize into regular patterns or crystalline structures to minimize overlaps, effectively navigating through the exclusion zones. These patterns exhibit long-range positional correlations, akin to those in traditional crystalline solids.

In \cite{helbing2}, the authors proposed a non-equilibrium thermodynamic potential for a system of repulsively interacting, oppositely moving particles and concluded that the most stable configuration minimizes interaction rates, achieved when individuals organize into uniform lanes. Another study by the same authors introduced the phenomenon of \textit{freezing-by-heating} \cite{freeze_by_heat}, where crystalline states emerge in repulsive two-particle systems as the system's total energy increases, analogous to pedestrian behavior under highly uncertain or panicked conditions. Our model captures both phases under different noise regimes, emphasizing the potential for developing a hydrodynamic framework that unifies the universal characteristics of such phases in repulsive self-propelled particle systems. Furthermore, it is worth investigating if constraining particles to hard-core interactions and varying the driving speed could reproduce the traveling and resting crystal phases observed in \cite{menzel2}.



\section*{Data Availability}
A python program of the simulation along with the data accompanying the results are available in  \href{https://github.com/bisht-prachi/repulsive-active-particles}{this repository}. Animation of the dynamical evolution of a system of $N=10,000$ active particles each with a radius of influence $R=0.1L$ in a domain of size $L=100$ is available as Supplementary data file \href{https://drive.google.com/file/d/1bjr9AiMrZ25XgHEz9LWlW79PcYmrS0kp/view?usp=share_link}{here} 
(or see Fig. \ref{fig:evolution}).

\bibliography{apssamp}%

\newpage

\appendix

 \begin{figure*}[htp]
    \begin{subfigure}{.3\textwidth}
        \centering
        \includegraphics[scale=0.25]{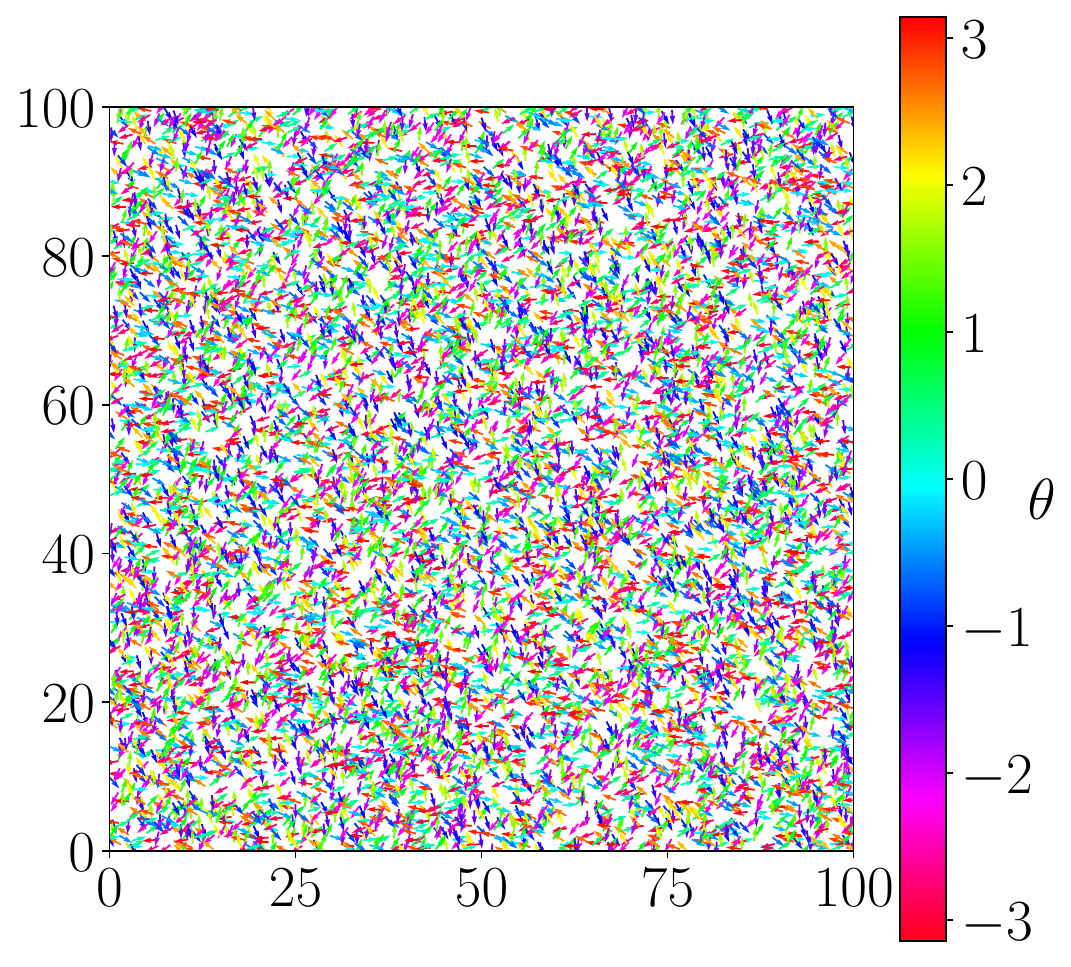}  
        \caption{t = 0}
    \end{subfigure}
    \begin{subfigure}{.3\textwidth}
        \centering
        \includegraphics[scale=0.25]{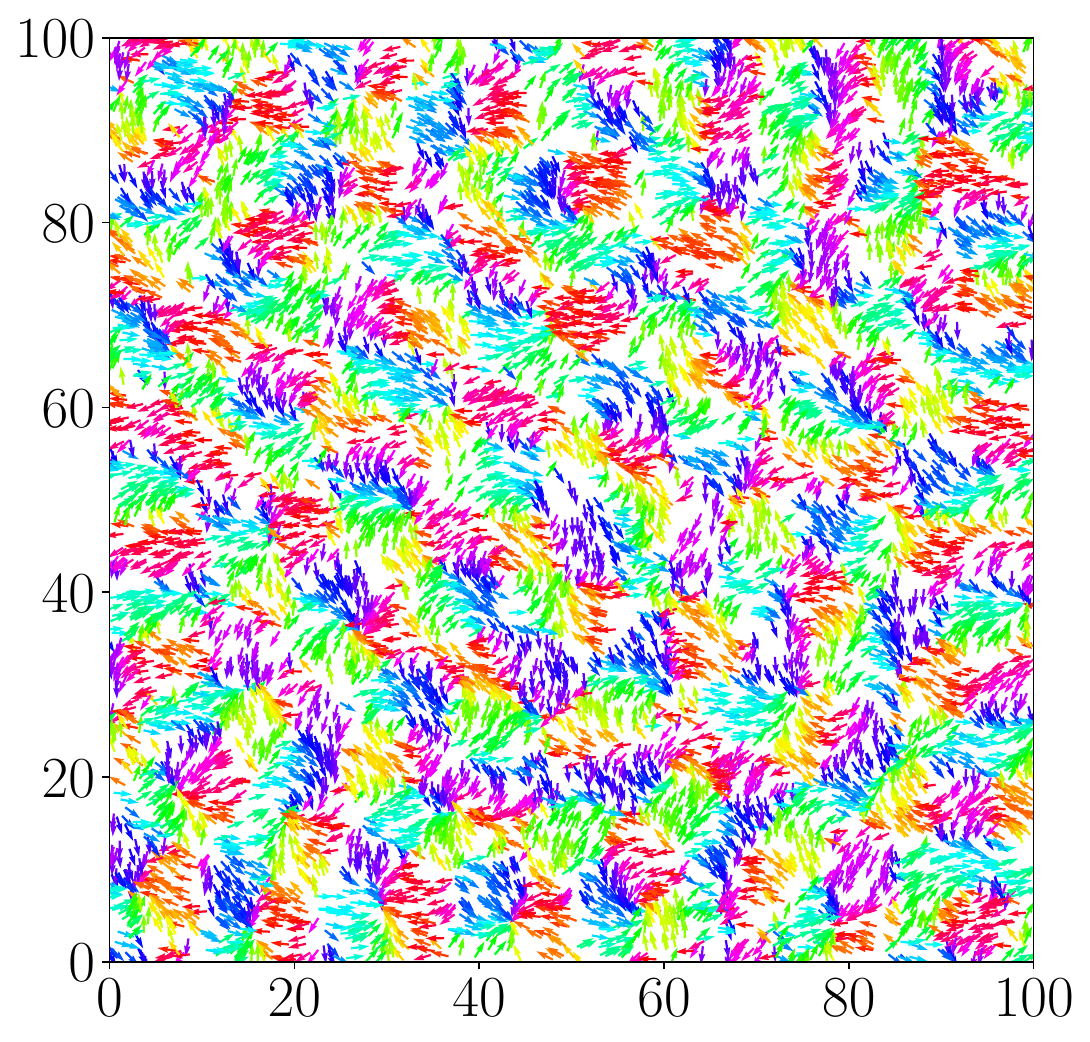}    
        \caption{t = 2}
    \end{subfigure}
    \begin{subfigure}{.3\textwidth}
        \centering
        \includegraphics[scale=0.25]{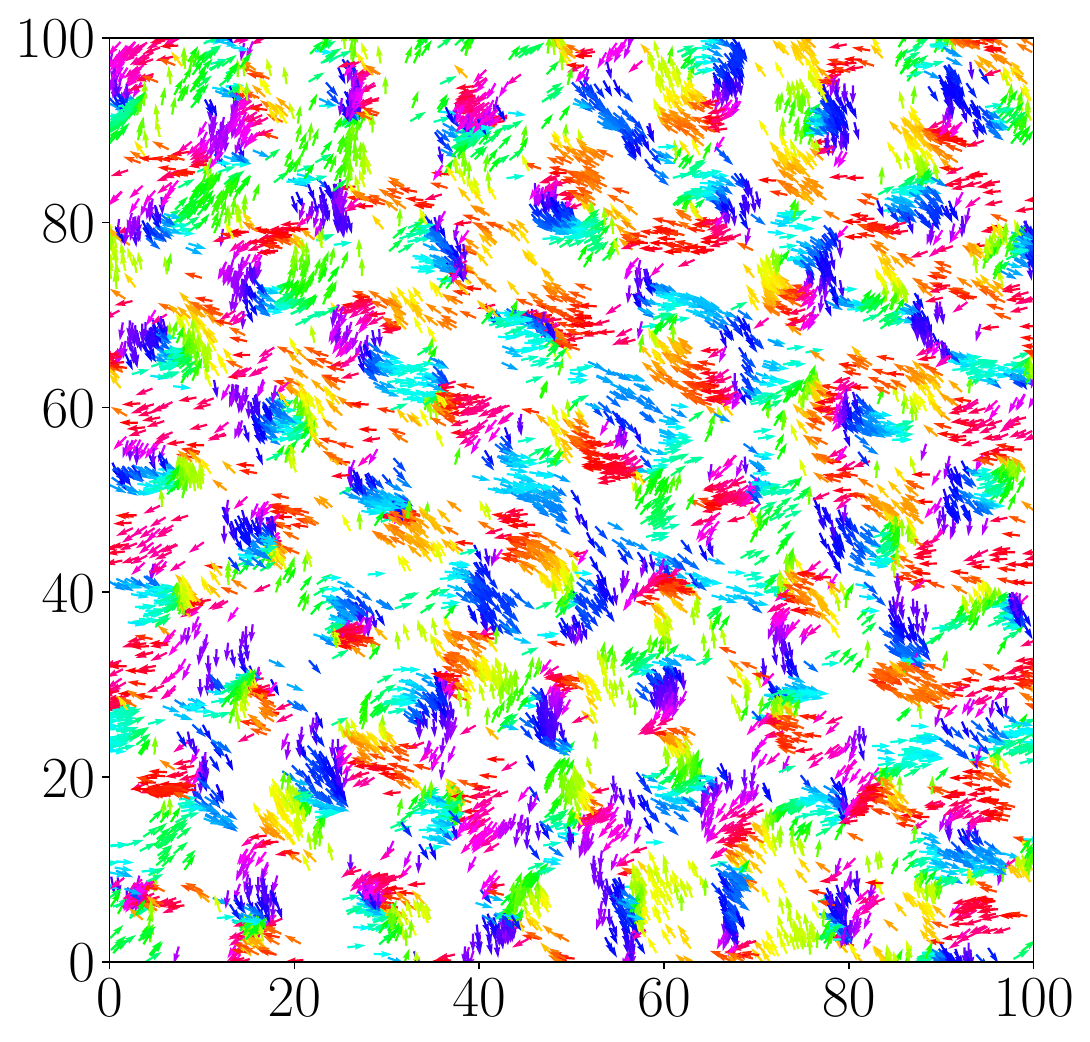}    
        \caption{t = 5}
    \end{subfigure}

    \begin{subfigure}{.3\textwidth}
        \centering
        \includegraphics[scale=0.25]{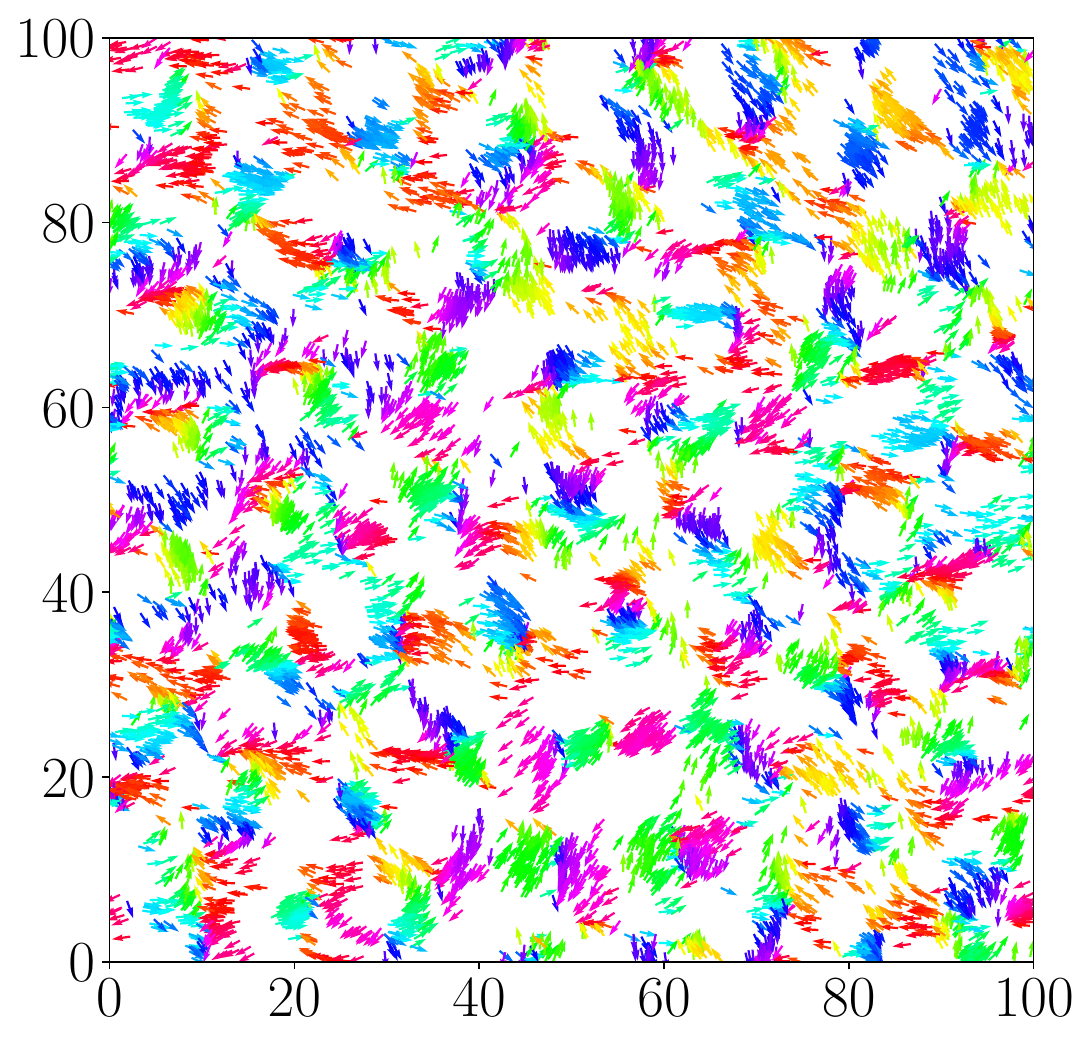}    
        \caption{t = 10}
    \end{subfigure}
    \begin{subfigure}{.3\textwidth}
        \centering
        \includegraphics[scale=0.25]{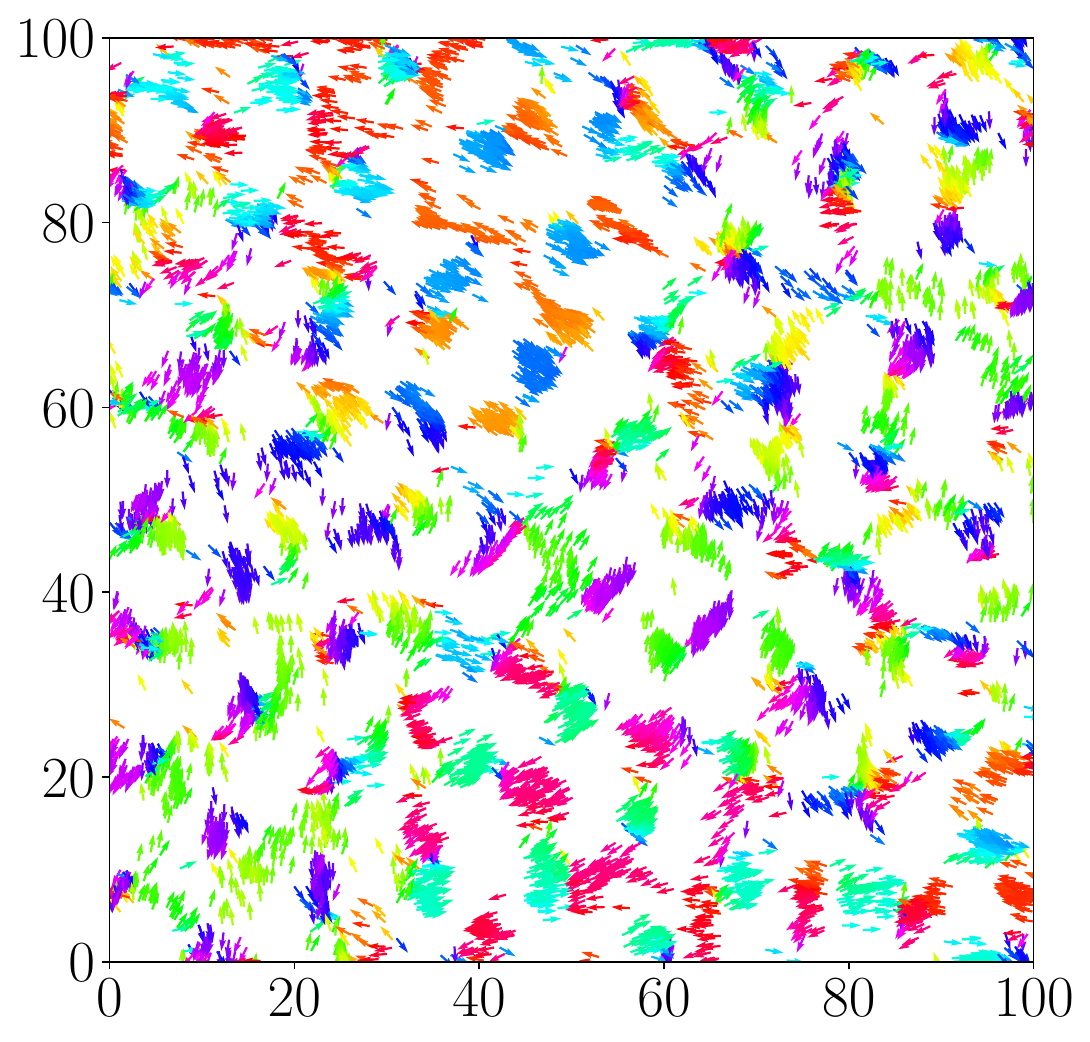}    
        \caption{t = 50}
    \end{subfigure}
    \begin{subfigure}{.3\textwidth}
        \centering
        \includegraphics[scale=0.25]{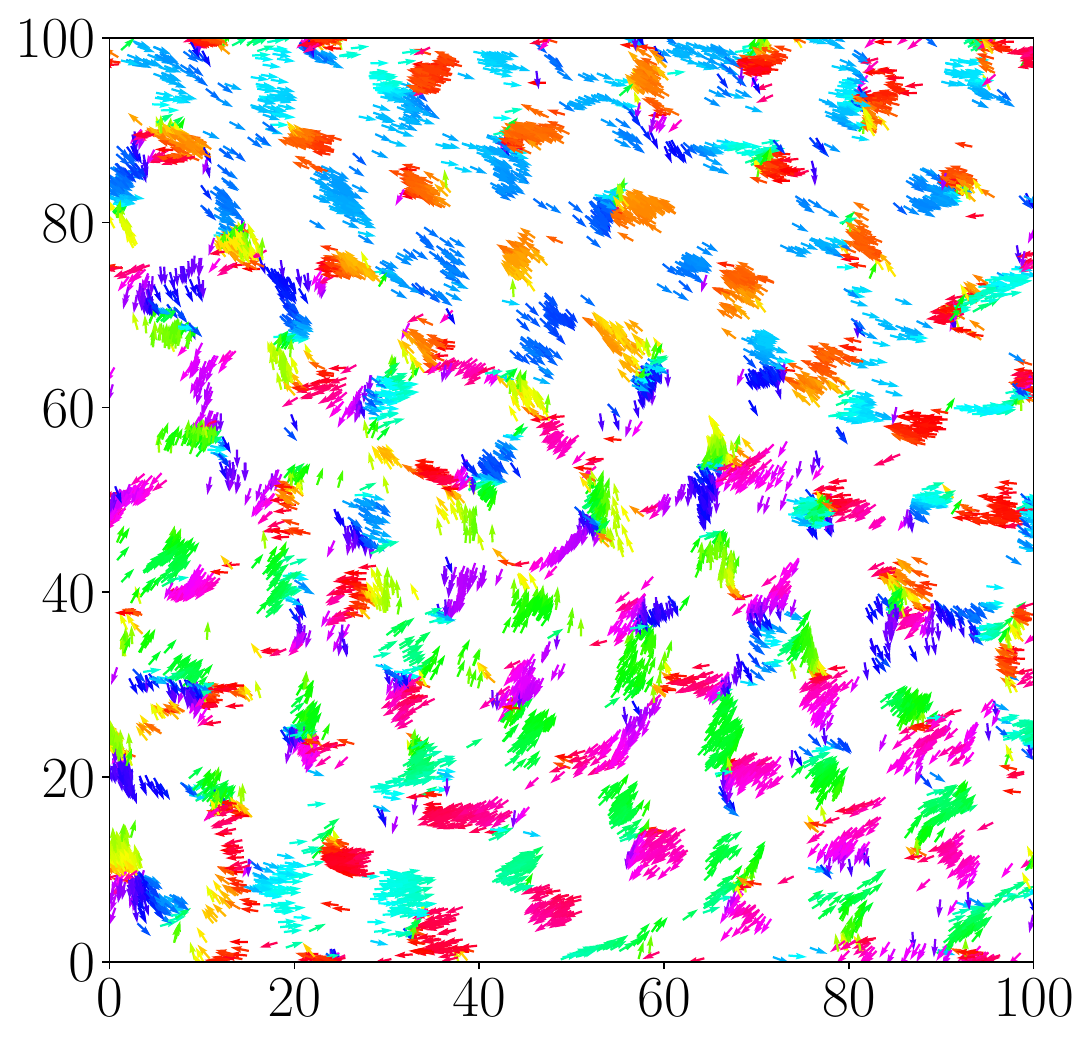}    
        \caption{t = 100}
    \end{subfigure}

    \begin{subfigure}{.3\textwidth}
        \centering
        \includegraphics[scale=0.25]{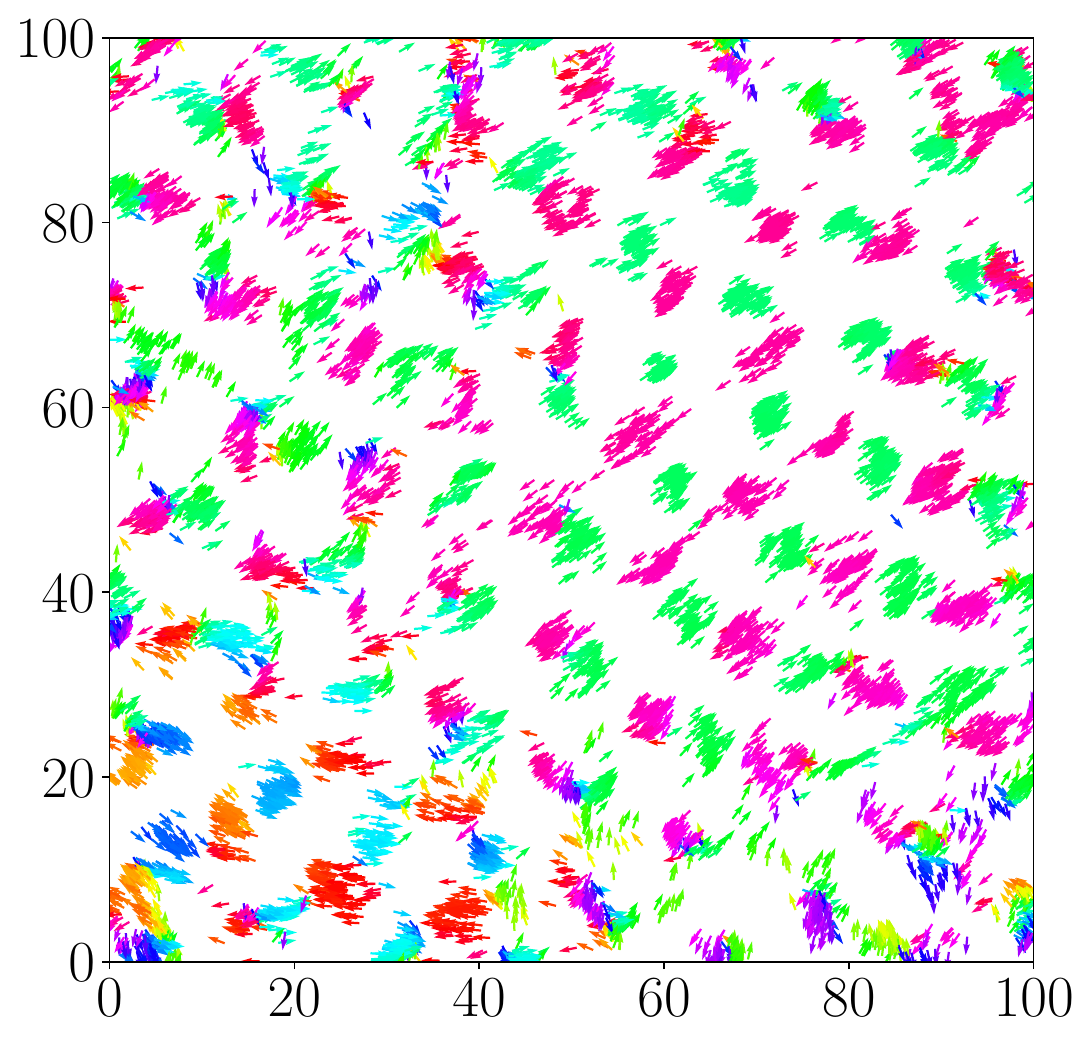}    
        \caption{t = 250}
    \end{subfigure}
    \begin{subfigure}{.3\textwidth}
        \centering
        \includegraphics[scale=0.25]{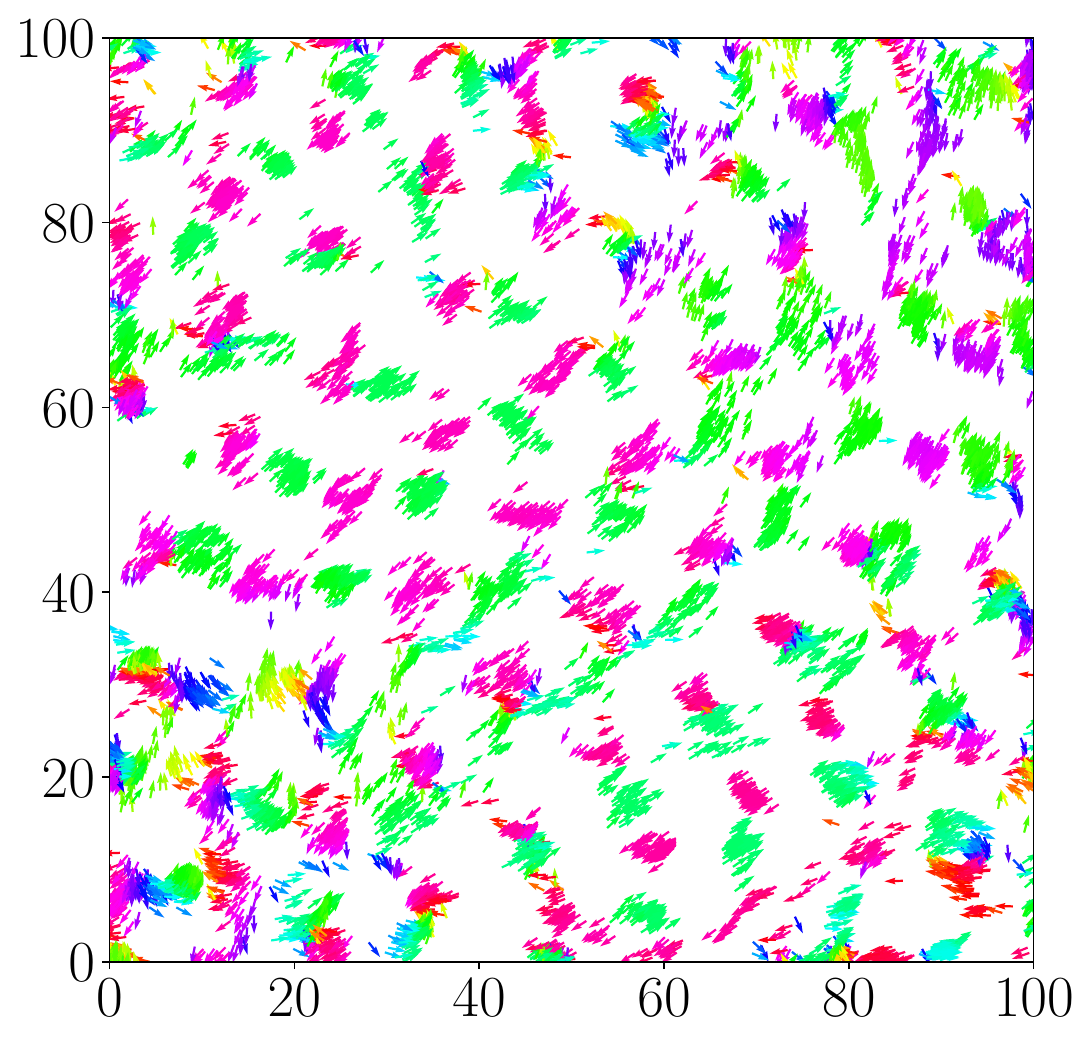}    
        \caption{t = 500}
    \end{subfigure}
    \begin{subfigure}{.3\textwidth}
        \centering
        \includegraphics[scale=0.25]{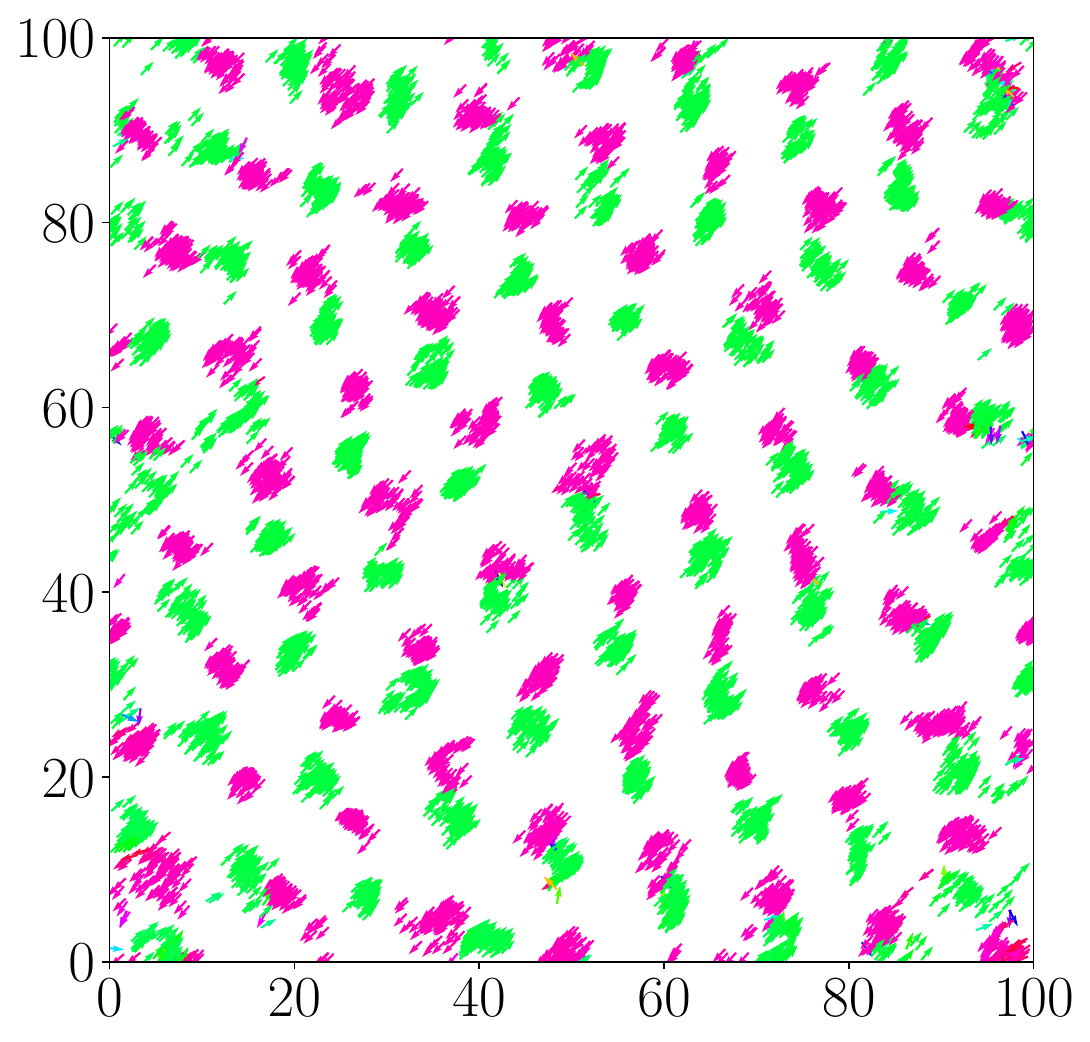}    
        \caption{t = 1500}
    \end{subfigure}
    
    \caption{Evolution snapshots of a system with $L=100, N=10,000, R=0.1L, \eta=0$ at simulation times as stated. 
        We see the formation of swirling vortices of particles right at the onset of the simulation. These vortices subsequently break down into clusters of particles with finite sizes which henceforth travel as one unit. There is a constant flux of particles in and out of these clusters; clusters break apart and reform all throughout the system (t=50). As time progresses (t=100), we observe large domains of clusters with opposing polarities co-existing, along partially developed lanes. Eventually, one dominant polarity prevails while smaller domains of other polarities get converted (t=250-500), resulting in a globally stable bi-lane system in the steady state (t=1500).}
    \label{fig:evolution}
 \end{figure*}

\newpage

\appendix

 \begin{figure*}[htp]
    \begin{subfigure}{.45\textwidth}
        \centering
        \includegraphics[scale=0.35]{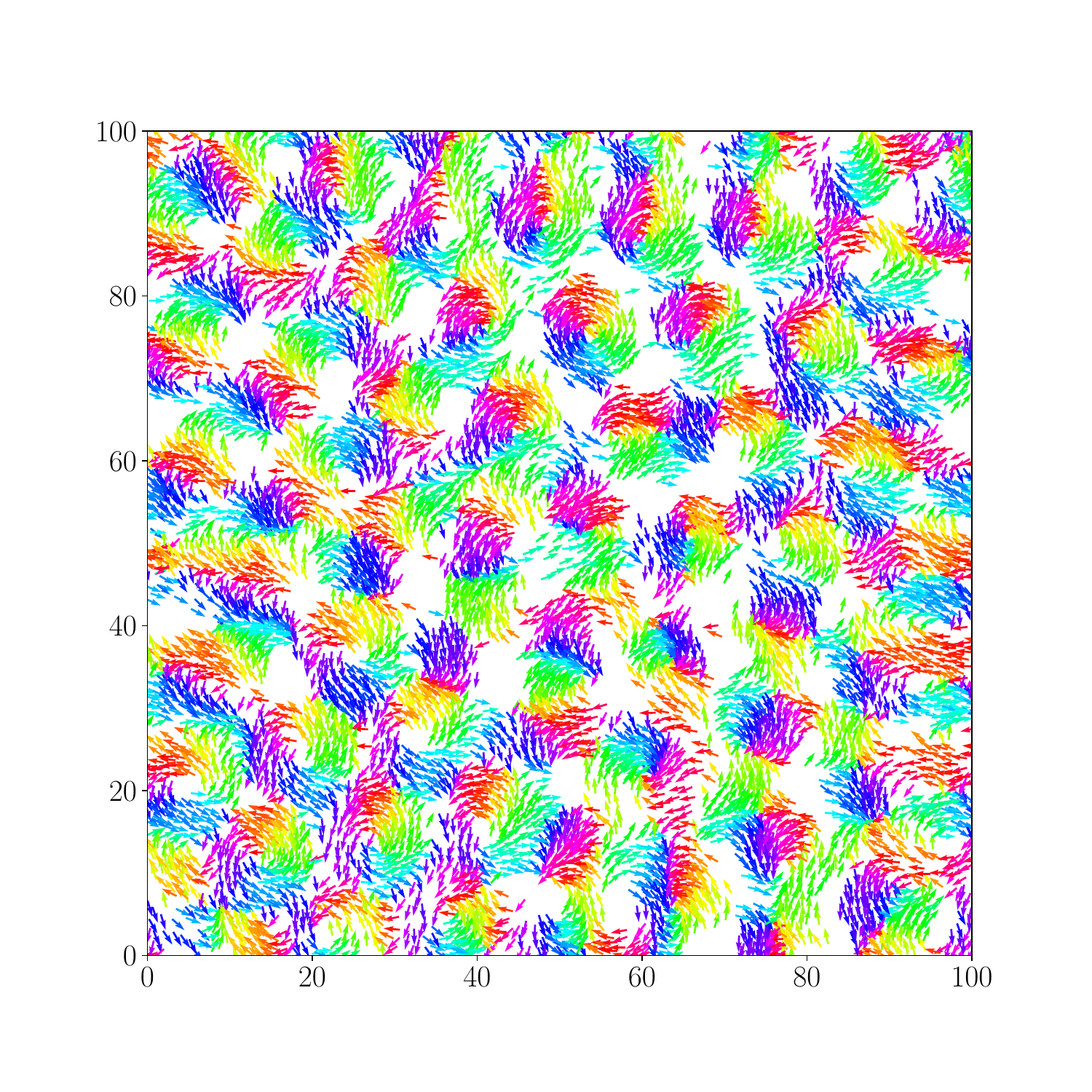}    
        \caption{t = 1500}
    \end{subfigure}

    \caption{Steady-state snapshot of the same simulation in Fig. \ref{fig:evolution} but under a hard-core constraint where minimum interparticle distance is set to an arbitrary constant (here, 1 unit).}
    \label{fig:evolution_hardcore}
 \end{figure*}
 
\end{document}